\documentclass[12pt]{article}
\input{psfig.tex}

\begin{document}

\title{Can the Universe escape eternal acceleration? }
\author{John D. Barrow$^1$, Rachel Bean$^2$, and Jo\~{a}o Magueijo$^2$ \\
$^1$DAMTP, Centre for Mathematical Sciences, \\
Cambridge University, Wilberforce Road, \\
Cambridge CB3 0WA.\\
$^2$Theoretical Physics, The Blackett Laboratory, \\
Imperial College, Prince Consort Rd.,\\
London SW7 2BZ}
\date{}
\maketitle

{\bf Recent astronomical observations of distant supernovae
light-curves, \cite
{super}-\cite{super3}, suggest that the expansion of the universe has
recently begun to accelerate. Acceleration is created
by an anti-gravitational repulsive stress, like that produced by a positive
cosmological constant, or universal vacuum energy. It
creates a rather bleak eschatological picture. An
ever-expanding universe's future appears to be increasingly dominated by its
constant vacuum energy. A universe doomed to accelerate forever will produce
a state of growing uniformity and cosmic loneliness.
Structures participating in the cosmological
expansion will ultimately leave each others' horizons and
information-processing must eventually die out\cite{BT}. Here, we examine
whether this picture is the only interpretation of the observations.
We find that in many well-motivated scenarios the observed spell of vacuum
domination is only a transient phenomenon. Soon after acceleration starts,
the vacuum energy's anti-gravitational properties are reversed,
and a matter-dominated
decelerating cosmic expansion resumes. Thus, contrary to general
expectations, once an accelerating universe does not mean always an
accelerating universe.}

The observed cosmic acceleration, if due to
vacuum repulsive stresses,  is very puzzling. In contrast to
conventional forms of matter, like pressureless dust or radiation, the
cosmological energy densities contributed by these repulsive stresses are
not diluted by expansion: they remain constant. Consequently, the vacuum
energy will eventually exceed the densities of all other forms of matter in
an ever-expanding universe, causing its expansion to accelerate. Yet, in
order to explain why the vacuum energy comes to dominate other matter only
close to the present time, $15Gyr$ after the expansion began, we must assume
it starts out with a density that is $10^{120}$ times smaller than
that of other forms of matter, for no known reason. A possible
explanation is the 'quintessence' picture, in which a scalar matter
field is present in the universe and can display repulsive gravitational
behaviour late in the universe's history, despite remaining innocuous for
most of its life \cite{quint}. In the most extreme case, its influence can
become identical to the presence of a positive cosmological constant with
constant energy density but the possible time-variation permitted in the
evolution of its density allows acceleration to appear at late times
following more natural initial conditions. We shall show
that  in  well-motivated
scenarios of this type the observed spell of vacuum domination is only
a transient phenomenon.

\texttt{\ }We consider a homogeneous and isotropic universe with zero
spatial curvature that contains two dominant forms of matter: a perfect
fluid with pressure $p$ and density $\rho $ linked by an equation of state $%
p=(\gamma -1)\rho $, with $\gamma $ constant, together with a hypothetical
scalar 'quintessence' field $\phi $ defined by its self-interaction
potential $V(\phi )$. All variables can depend only on the cosmic co-moving
proper time $t$. The cosmological expansion scale factor is $a(t)$ and the
Hubble expansion rate is defined by $H\equiv \dot a/a.$ Separate
conservation of mass-energy for the two matter fields as the universe
expands requires $\rho \propto a^{-3\gamma }$, while the scalar field, $\phi
$, obeys
\begin{equation}
\ddot \phi +3{H}\dot \phi ={\frac{\partial V}{\partial \phi },}  \label{A}
\end{equation}
where overdots represent time derivatives. The densities of the perfect
fluid, $\rho $, and the $\phi $ field, $\rho _\phi $, drive the expansion, so
\begin{equation}
3H^2=\rho +\rho _\phi ,  \label{B}
\end{equation}
where the scalar-field's energy density is the sum of its kinetic and
potential parts, $\rho _\phi \equiv \frac 12\dot \phi ^2+V.$ We use Planck
units defined by $\hbar =c=(8\pi G)^{-1/2}=1.$

If we choose a monotonic potential of the form $V(\phi )=Ae^{-\lambda
\phi }$, with $A$ and $\lambda >{\sqrt{2}}$ constant parameters, then $V$
need not increasingly dominate the right-hand side of eq. (\ref{B}) at large
$a$. Instead, $\rho _\phi $ evolves in direct proportion to $\rho $ $%
\propto t^{-2}$ during an era of radiation ($\gamma =4/3$) or matter ($\gamma
=1$) domination. Hence
$H=2/3\gamma t$ and the fraction of the total density in
the form of quintessence during this 'scaling' regime depends only on $%
\lambda :$
\[
\Omega _\phi \equiv \frac{\rho _\phi }{\rho _\phi +\rho }=\frac{\rho _\phi }{%
3H^2}=\frac{3\gamma }{\lambda ^2}.
\]

This unusual type of behaviour shows how vacuum domination can be postponed
for a long period since the universe's expansion began. However, eventually,
the expansion departs from this special scaling behaviour and the potential
of the $\phi $ field comes to dominate eq. (\ref{B}), contributing an almost
constant value of $\rho _\phi $ and accelerating expansion. Eventually, the
potential energy $V$ supersedes the kinetic energy,$\frac 12\dot \phi ^2,$
making quintessence behave like an effective cosmological constant. In the
standard scenarios the early scaling behaviour is a robust feature of
quintessence models, but subsequent late-time acceleration can only be
achieved with a large degree of fine tuning of the defining parameters -
either in the initial conditions or in the parameters of the potential.

Albrecht and Skordis \cite{alb} have proposed a particularly attractive
model of quintessence. It is driven by a potential which introduces a small
minimum to the exponential potential:
\begin{equation}
V(\phi )=e^{-\lambda \phi }{\left( A+(\phi -\phi _0)^2\right) .}  \label{C}
\end{equation}
Unlike previous quintessence models, late-time acceleration is achieved
without fine tuning of the initial conditions. The authors argue that such
potentials arise naturally in the low-energy limit of $M$-theory; the
constant parameters, $A$ and $\phi _0$, in the potential take values of
order $1$ in Planck units, so there is also no fine tuning of the potential.
They show that, regardless of the initial conditions, $\rho _\phi $ scales,
with $\rho \propto \rho _\phi \propto t^{-2}$ during the radiation and
matter eras, but leads to permanent vacuum domination and accelerated
expansion after a time which can be close to the present. Acceleration
begins when the field gets trapped in the local minimum of the potential at
$\phi =\phi _0+(1\pm \sqrt{1-\lambda ^2A})/\lambda $,
which is created by the quadratic factor in eq. (\ref{C}) when $1\geq
\lambda ^2A$.
Once the field gets stuck in the false vacuum its kinetic energy disappears (%
$\phi \approx $ constant), and the ensuing dominance of $\rho +\rho _\phi $
by an almost constant value of the potential value triggers a period
of accelerated expansion that never ends. The probability for quantum
tunnelling through the barrier is negligible \cite{joch}.

\begin{figure}
\centerline{\psfig{file=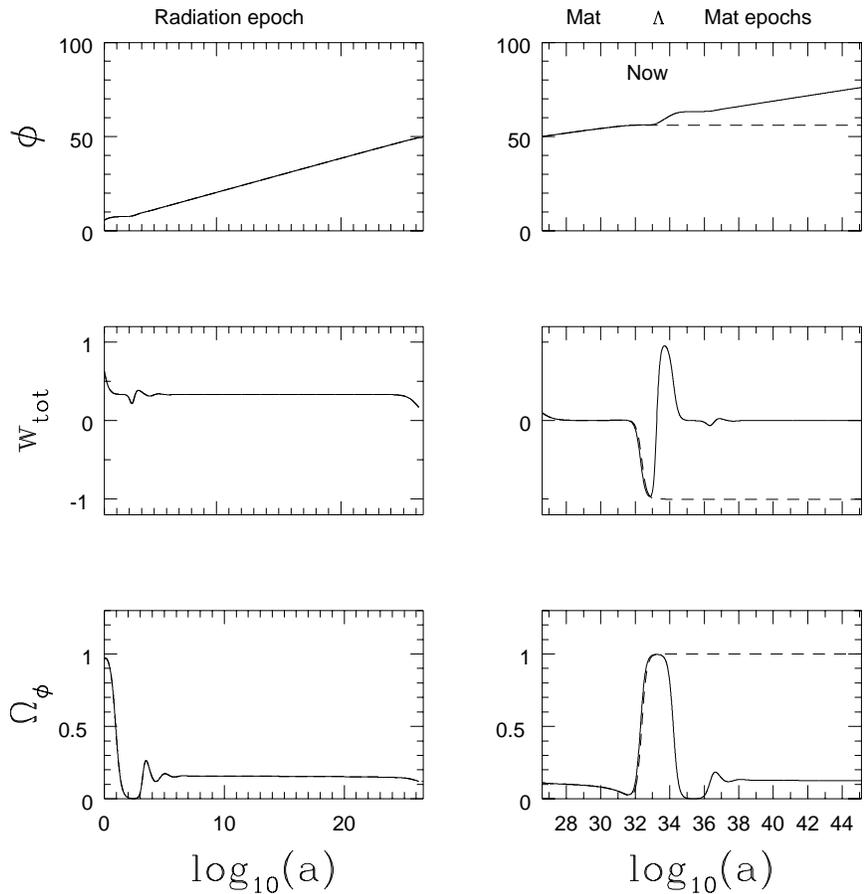,width=12 cm,angle=0}}
\caption{Permanent (dashed) and transient (solid)
vacuum domination (with $\lambda =4$, and $A\lambda ^2=0.5, 1$ respectively).
The early scaling behaviour is identical, but in one case the field gets
stuck in the local potential minimum, leading to permanent vacuum
domination, while in the other case the field resumes rolling down the
potential, and a second period of matter-dominated expansion follows the
temporary vacuum domination.}
\label{fig1}
\end{figure}

We have found that this type of behaviour is by no means generic. In Fig.~%
\ref{fig1} we plot the evolution of $\Omega _\phi$
and a measure of the effective equation of state of the total
matter content of the universe $w_{tot}=p_{tot}/\rho _{tot}$, where the
total density is $\rho _{tot}=$ $\rho +\rho _\phi $ and the total pressure is
$p_{tot}=(\gamma -1)\rho +\frac 12\dot{\phi}^2-V$.
We show $\Omega _\phi $ and $w_{tot}$ in two qualitatively distinct cases,
created by equally plausible values for potential parameters. Vacuum
domination occurs when $\Omega _\phi >1/2$, and accelerated expansion when $%
w_{tot}<-1/3$. The dashed line traces the case found by Albrecht and
Skordis, but the solid line shows the general behaviour. In both cases there
is scaling of the densities early in the expansion history: with $w_\phi =1/3
$ and $\Omega _\phi \approx 4/\lambda ^2$ in the radiation era, followed by $%
w_\phi =0$ and $\Omega _\phi \approx 3/\lambda ^2$ in the matter era . The
expansion decelerates in both scaling eras. But, then, in both cases, as $%
\phi \rightarrow \phi _0$, vacuum domination and accelerated expansion is
triggered. Whereas in the first case this phenomenon is permanent, in the
latter it is ephemeral. In the latter case, $\phi $ soon continues rolling
down the potential, and the universe resumes scaling evolution with $\rho $ $%
\propto \rho _\phi $ $\propto t^{-2}$ and another matter-dominated era
ensues. The spectre of never-ending vacuum domination has been lifted.

Transient vacuum domination arises in two ways. When $A\lambda ^2<1,$ the $%
\phi $ field arrives at the local minimum with enough kinetic energy to roll
over the barrier and resume descending the exponential part of the potential
where $\phi >>\phi _0$. This kinetic energy is determined by the scaling
regime, and so by parameters of the potential and not initial conditions.
Another instance of transient vacuum domination is the whole region $%
A\lambda ^2>1$. As $A$ increases towards $\lambda ^{-2}$, the potential
loses its local minimum, and flattens out into a point of inflexion. This is
sufficient to trigger accelerated expansion temporarily, but the field never
stops rolling down the potential, and matter-dominated scaling evolution
with $a(t)\propto t^{2/3}$ is soon resumed.

\begin{figure}
\centerline{\psfig{file=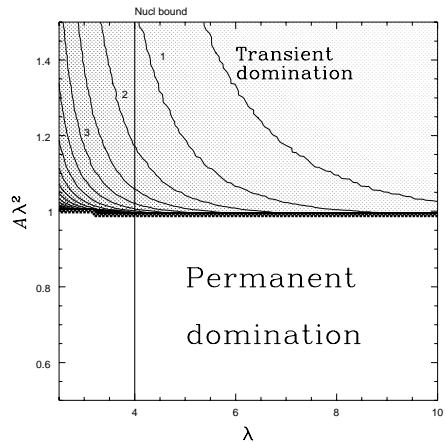,width=6 cm,angle=0}}
\caption{The phase space of the Albrecht-Skordis
theory. A critical line divides permanent from transient vacuum domination.
Above this line we have plotted contours for the number of e-foldings during
which the Universe accelerates. The line $\lambda\approx 4$ indicates
the nucleosynthesis bound ($\lambda>4$).}
\label{fig2}
\end{figure}

Further scrutiny of the theory reveals a critical dividing line separating
permanent and transient vacuum domination. If we impose on the parameter
space $\{A,\phi _0,\lambda \}$ the condition that accelerated expansion be
occurring by the recent past, as observations imply, then we require $%
\lambda \phi _0\approx 280$. This leaves us with two degrees of freedom,
which we parametrise by $\lambda $ and $A\lambda ^2$. The behaviour of the
Universe in this parameter space is shown in Fig.~\ref{fig2}. The unshaded
area represents the regions of permanent vacuum domination. Outside this
area we have plotted contours giving the number of e-foldings of accelerated
expansion, defined by $N=log(a_f/a_i)$, where $a_f$ and $a_i$ are the values
of the expansion scale factor at the end and start of acceleration,
respectively.

Our investigations have revealed a new type of cosmological evolution. It is
possible for the universe to exit from a period of accelerated expansion and
resume decelerated expansion. Moreover, for the well-motivated family of
Albrecht-Skordis potentials this is the most likely form of evolution,
rather than a state of continuing acceleration. Models of this type create a
different theoretical framework in which to interpret the observation
results of \cite{super}-\cite{super3} and change their consequences for
theories of cosmic structure formation. Other escape routes from
acceleration, like introducing a quintessence field that ultimately decays
into matter and radiation after producing acceleration, are at present
entirely ad hoc. We can also imagine a richer cosmological structure if we
admit the possibility of a random variation of $V(\phi )$ around the
universe, in the spirit of chaotic inflationary universes. The condition for
transient accelerated expansion will arise quasi-randomly around the
universe as a set of parameters like $\{A,\phi _0,\lambda \}$ varies
stochastically. This will give rise to some environments displaying
transient acceleration at different times in their history, in which galaxy
formation and clustering can continue to develop for longer than in the
rarer, perpetually-accelerating, regions. The ability of expanding regions
to recover from the onset of accelerated vacuum-dominated expansion may be
an important factor in the global evolution of regions of an infinite
universe containing long-lived stars and galaxies.


\end{document}